\documentclass[conference]{IEEEtran}
\ifCLASSINFOpdf
   \usepackage[pdftex]{graphicx}
\else
\fi
\begin{document}
%
\title{Evolution and Innovation in 5G Cellular Communication
System and Beyond: A Study
}

\author{\IEEEauthorblockN{Rehman Talukdar and Mridul Saikia}
\IEEEauthorblockA{Department of Information Technology \\
North-Eastern Hill University\\
Shillong 793 022 India}
}


%


\maketitle

\begin{abstract}
Since the last few years there has been a phenomenal growth in the wireless industry. Widespread wireless technologies,
increasing variety of user-friendly and multimedia-enabled terminals and wider availability of open source tools for content generation has lead encouraged user-centric networks resulting in a need for efficient network design. The objective of this paper is comprehensive study related to 5G technology of mobile communication. Existing research work in mobile communication is related to 5G technology. The major contribution of this study is the key provisions of 5G (Fifth Generation) technology of mobile communication, which is seen as consumer oriented. In 5G technology, the mobile consumer has given utmost priority compared to others. In this context, the existing and highly demanded technologies for 5G technologies has beed studied extensively. Open challenges are highlighted for researcher for further study of the emerging 5G systems.
\end{abstract}

\section{Introduction}
\subsection{5G Technology}
Mobile and wireless networks have made tremendous growth in the last fifteen years. Nowadays many mobile phones have also a WLAN adapter. One may suppose that near soon many mobile phones will have WiMAX adapter too, besides their 3G, 2G, WLAN, Bluetooth etc. adapters. Today 3G mobile systems are on the ground providing IP connectivity for real-time and non-real-time services. On the other side, there are many wireless technologies that have proven to be important, with the most important ones being 802.11 Wireless Local Area Networks (WLAN) and 802.16 Wireless Metropolitan Area Networks (WMAN), as well as ad-hoc Wireless Personal Area Network (WPAN) and wireless networks for digital TV and radio broadcast. Then, the concepts of 4G is already much discussed and it is almost certain that 4G will include several standards under a common umbrella, similarly to 3G, but with IEEE 802.xx wireless mobile networks included from the beginning. The main contribution of this paper is definition of 5G (Fifth Generation) mobile network concept, which is seen as user-centric concept instead of operator-centric as in 3G or service-centric concept as seen for 4G. In the proposed concept the mobile user is on the top of all. The 5G terminals will have software defined radios and modulation scheme as well as new error-control schemes can be downloaded from the Internet on the run. The development is seen towards the user terminals as a focus of the5G mobile networks. The terminals will have access to different wireless technologies at the same time and the terminal should be able to combine different flows from different technologies. Each network will be responsible for handling user-mobility, while the terminal will make the final choice among different wireless/mobile access network providers for a given service. The paper also proposes intelligent Internet phone concept where the mobile phone can choose the best connections by selected constraints and dynamically change them during a single end-to-end connection. The proposal in this paper is fundamental shift in the mobile networking philosophy compared to existing 3G and near-soon 4G mobile technologies, and this concept is called here – the 5G.
\par
5G technology has changed the means to use cell phones within very high bandwidth. User never experienced ever before such a high value technology. Nowadays mobile users have much awareness of the cell phone (mobile) technology. The 5G technologies include all type of advanced features which makes 5G technology most powerful and in huge demand in near future.
\par
The gigantic array of innovative technology being built into new cell phones is stunning. 5G technologies which are on hand held phone offering more power and features than at least 1000 lunar modules. A user can also hook their 5G technology cell phone with their Laptop to get broadband internet access. 5G technology including camera, MP3 recording, video player, large phone memory, dialling speed, audio player and much more you never imagine. For children rocking fun Bluetooth technology and Pico nets has become in market.
\begin{figure*}[tb]
  \caption{Advanced Heterogeneous Network Architecture}
  \centering
    \includegraphics[width=1.0\textwidth]{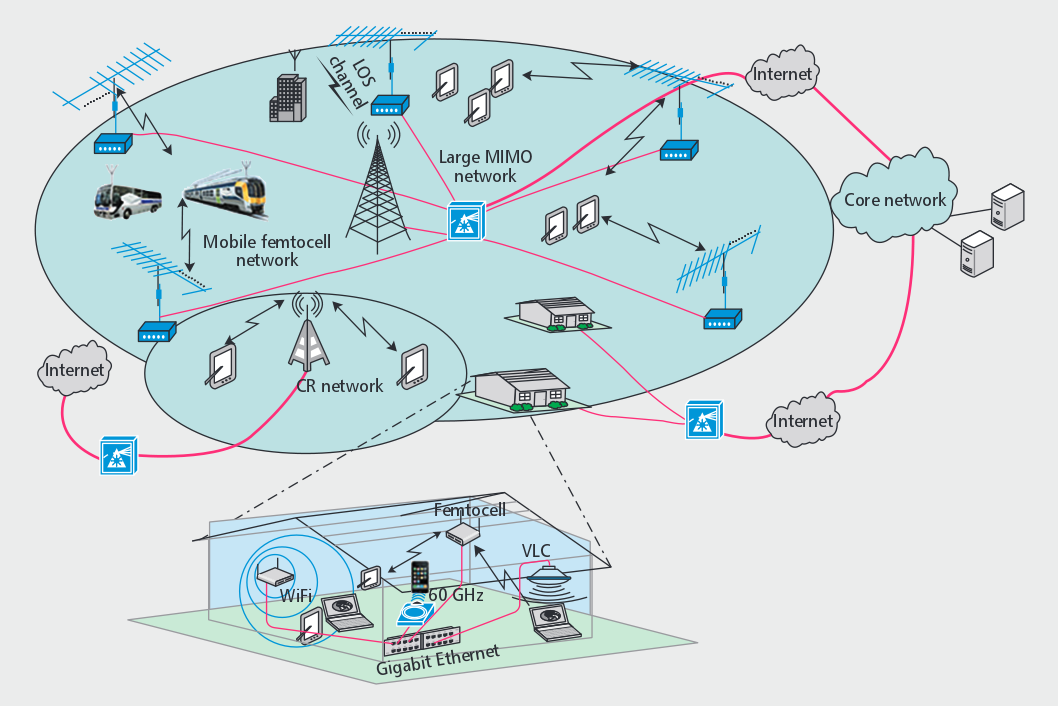}
\end{figure*}

\subsection{What 5G Technology Offers}
5G technology going to be a new mobile revolution in mobile market. Through 5G technology now you can use worldwide cellular phones and this technology also strike the china mobile market and a user being proficient to get access to Germany phone as a local phone. With the coming out of cell phone alike to PDA now your whole office in your finger tips or in your phone. 5G technology has extraordinary data capabilities and has ability to tie together unrestricted call volumes and infinite data broadcast within latest mobile operating system. 5G technology has a bright future because it can handle best technologies and offer priceless handset to their customers. May be in coming days 5G technology takes over the world market. 5G Technologies have an extraordinary capability to support Software and Consultancy. The Router and switch technology used in 5G network providing high connectivity. The 5G technology distributes internet access to nodes within the building and can be deployed with union of wired or wireless network connections. The current trend of 5G technology has a glowing future. The figure 1 below shows an example of a HetNet in a LTE-Advanced Network.
\begin{figure*}[tb]
  \caption{Comparison of All Generations of Mobile Technologies (1G - 5G).}
  \centering
    \includegraphics[width=1.0\textwidth]{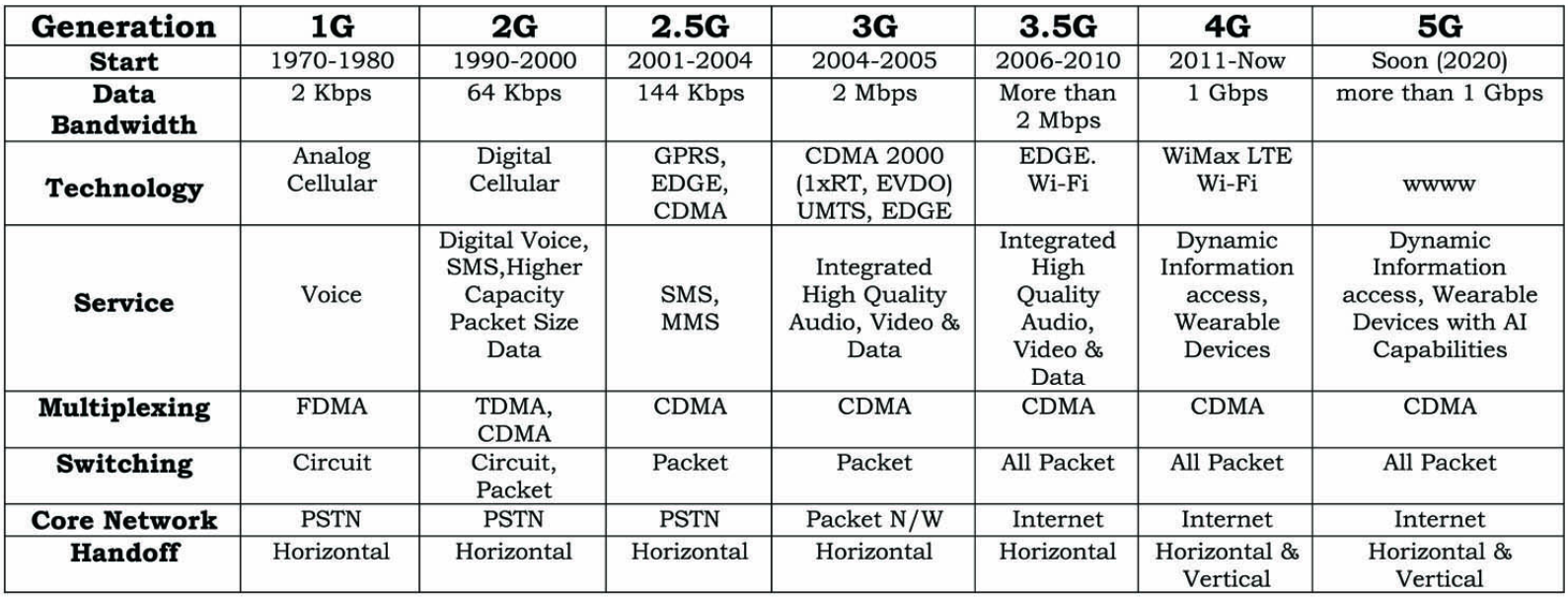}
\end{figure*}
\section{Visions and Requirements}
In January 2012 ITU approved the first release of the 4G global core standard (GCS). After the world radio communication conferences 2012 (WRC-12), telecommunication industry began to discuss the visions and requirements of 5G.
\par
5G is needed because of the explosive growth in video traffic, the acute shortage of spectrum, the growing need to minimise the energy requirements of web devices and network infrastructure and to cater to the insatiable desire
for higher data speed rates.
\par
For the customer, the difference between 4G and 5G technologies will be in higher speeds, lower battery consumption, better coverage, higher number of supported devices, lower infrastructure costs, higher versatility or higher reliability of communications.
\par
Mobile communication systems have been playing an important role in our life for more than 20 years, and it
will enter even more dimensions of the society due to technology improvement. 5G is expected to play even a
larger role in the year 2020 and beyond, and its social responsibilities and functions can be summarized as
following four aspects.
\section{Concepts for 5G Mobile Networks}
The 5G terminals will have software defined radios and modulation schemes as well as new error-control schemes that can be downloaded from the Internet. The development is seen towards the user terminals as a focus of the 5G mobile networks. The terminals will have access to different wireless technologies at the same time and the terminal should be able to combine different flows from different technologies. The vertical handovers should be avoided, because they are not feasible in a case when there are many technologies and many operators and service providers. In 5G, each network will be responsible for handling user-mobility, while the terminal will make the final choice among different wireless/mobile access network providers for a given service. Such choice will be based on open intelligent middleware in the mobile phone.
\subsection{Physical Layer}
Physical and Medium Access Control layers i.e. OSI layer 1 and OSI layer 2, define the wireless technology. For these two layers the 5G mobile networks is likely to be based on Open Wireless Architecture.

\subsection{Network Layer}
The network layer will be IP (Internet Protocol), because there is no competition today on this level. The IPv4 (version 4) is worldwide spread and it has several problems such as  limited address space and has no real possibility for QoS support per flow. These issues are solved in IPv6, but traded with significantly bigger packet header. Then, mobility still remains a problem. There is Mobile IP standard on one side as well as many micro-mobility solutions (e.g., Cellular IP, HAWAII etc.). All mobile networks will use Mobile IP in 5G, and each mobile terminal will be FA (Foreign Agent), keeping the CoA (Care of Address) mapping between its fixed IPv6 address and CoA address for the current wireless network. However, a mobile can be attached to several mobile or wireless networks at the same time. In such case, it will maintain different IP addresses for each of the radio interfaces, while each of these IP addresses will be CoA address for the FA placed in the mobile Phone. The fixed IPv6 will be implemented in the mobile phone by 5G phone manufactures.

\par
The 5G mobile phone shall maintain virtual multi-wireless network environment. For this purpose there should be separation of network layer into two sub-layers in 5G mobiles (Fig. ) i.e.: Lower network layer (for each interface) and Upper network layer (for the mobile terminal). This is due to the initial design of the Internet, where all the routing is based on IP addresses which should be different in each IP network world wide. The middleware between the Upper and Lower network layers (Fig. 3) shall maintain address translation from Upper network address (IPv6) to different Lower network IP addresses (IPv4 or IPv6), and vice versa.

\subsection{Open Transport Protocol (OTA) Layer}
The mobile and wireless networks differ from wired networks regarding the transport layer. In all TCP versions the assumption is that lost segments are due to network congestion, while in wireless networks losses may occur due to higher bit error ratio in the radio interface. Therefore, TCP modifications and adaptation are proposed for the mobile and wireless networks, which retransmit the lost or damaged TCP segments over the wireless link only. For 5G mobile terminals will be suitable to have transport layer that is possible to be downloaded and installed. Such mobiles shall have the possibility to download (e.g., TCP, RTP etc. or new transport  protocol) version which is targeted to a specific wireless technology installed at the base stations. This is called here Open Transport Protocol - OTP.

\subsection{Application Layer}
     Regarding the applications, the ultimate request from the 5G mobile terminal is to provide intelligent QoS management over variety of networks. Today, in mobile phones the users manually select the wireless interface for particular Internet service without having the possibility to use QoS history to select the best wireless connection for a given service. The 5G phone shall provide possibility for service quality testing and storage of measurement information in information databases in the mobile terminal. The QoS parameters, such as delay, jitter, losses, bandwidth, reliability, will be stored in a database in the 5G mobile phone with aim to be used by intelligent algorithms running in the mobile terminal as system processes, which at the end shall provide the best wireless connection upon required QoS and personal cost constraints.

\section{Conclusion and Future Enhancement}
\subsection{Conclusion}
     In this paper we have proposed 5G mobile phone concept, which is the main contribution of the paper. The 5G mobile phone is designed as an open platform on different layers, from physical layer up to the application. Currently, the ongoing work is on the modules that shall provide the best QoS and lowest cost for a given service using one or more than one wireless technology at the same time from the 5G mobile phone.
     A new revolution of 5G technology is about to begin because 5G technology going to give tough completion to normal computer and laptops whose marketplace value will be effected. There are lots of improvements from 1G, 2G, 3G, and 4G to 5G in the world of telecommunications. The new coming 5G technology is available in the market in affordable rates, high peak future and much reliability than its preceding technologies.
\subsection{Future Enhancement}
     5G network technology will open a new era in mobile communication technology. The 5G moble phones will have access to different wireless technologies at the same time and the terminal should be able to combine different flows from different technologies. 5G technology offer high resolution for crazy cell phone user. We can watch TV channels at HD clarity in our mobile phones without any interruption. The 5G mobile phones will be a tablet PC. Many mobile embedded technologies will evolve.


\begin{thebibliography}{1}
\bibitem{IEEEhowto:Toni}
Toni Janevski, “AAA System for PLMN-WLAN Internetworking”, Journal of    Communications and Networks (JCN), pp.192-206, Volume 7,Number 2, June 2005.
\bibitem{IEEEhowto:Janise}
Janise McNair, ang Zhu, “Vertical Handoffs in Fourth-Generation Multinetwork Environments”, IEEE Wireless Communications, June2004.
\bibitem{IEEEhowto:Toni}
Toni Janevski, “Traffic Analysis and Design of Wireless IP Networks”, Artech House Inc., Boston, USA, 400 p., May 2003.
\bibitem{IEEEhowto:Willie}
B. Sahoo, "Performance Comparison of Packet Scheduling Algorithms for Video Traffic in LTE Cellular Network", International Journal of Mobile Network Communications \& Telematics ( IJMNCT) Vol. 3, No.3, PP 09-18, June 2013
\bibitem{IEEEhowto:Suk}
Suk Yu Hui, Kai Hau Yeung, “Challenges in the Migration to 4G Mobile Systems”, IEEE Communications Magazine, December 2003.
\bibitem{IEEEhowto:Bria}
Bria, F. Gessler, O. Queseth, R. Stridth, M. Unbehaun, J.Wu, J.Zendler, “4-th Generation Wireless Infrastructures: Scenarios and Research Challenges”, IEEE Personal Communications, Vol. 8,                                                                   
\bibitem{IEEEhowto:Willie}
Willie W. Lu, “An Open Baseband Processing Architecture for Future Mobile Terminals Design”, IEEE Wireless Communications, April 2008.
\bibitem{IEEEhowto:Asvin}
Asvin Gohil, Hardik Modi, Shobhit K Patel, "5G Technology of Mobile Communication: A Survey", 2013 International Conference on Intelligent Systems and Signal Processing (ISSP).
\end{thebibliography}
\end{document}